\input mn

\pagerange{1}
\pubyear{1996}
\volume{000}

\begintopmatter  
\title{The metallicity distribution of G dwarfs in the solar neighbourhood}
\author{H. J. Rocha-Pinto and W. J. Maciel}
\affiliation{Instituto Astron\^omico e Geof\'{\i}sico, Av. Miguel Stefano 4200, 04301-904 
S\~ao Paulo, Brazil}

\shortauthor{H. J. Rocha-Pinto and W. J. Maciel}
\shorttitle{The metallicity distribution of G dwarfs}

\acceptedline{Accepted 1995 October 4. Received 1995 September 15; in 
original form 1995 May 30.}

\abstract {We derive a new metallicity distribution of G dwarfs in the 
solar neighbourhood, using $uvby$ photometry and up-to-date parallaxes. 
Our distribution comprises 287 G dwarfs within 25 pc from the Sun, and 
differs considerably from the classic solar neighbourhood distribution of 
Pagel \& Patchett and Pagel by having a prominent single 
peak around $\rm{[Fe/H]} = -0.20$ dex. The raw data are corrected for 
observational errors and cosmic scatter assuming a deviation $\sigma=0.1$. 
In order to obtain the true abundance distribution, we use 
the correction factors given by Sommer-Larsen, which take into 
account the stellar scale heights. 

The distribution confirms the G dwarf problem, that is, the paucity of 
metal-poor stars relative to the predictions of the simple model of 
chemical evolution. Another feature of this distribution, which was 
already apparent in previous ones, is the small number of metal-rich stars 
again in comparison with the simple model. Our results indicate 
that it is very difficult to fit the simple model to this 
distribution, even with the definition of an `effective yield'. A 
comparison with several models from the literature is made. We find 
that models with infall are the most appropriate to explain the new 
metallicity distribution. We also show that the metallicity distribution 
is compatible with a major era of star formation occurring 5 to 8 Gyr 
ago, similar to results found by several authors.}


\maketitle  

\section{Introduction}

\tx 
The so-called G dwarf problem was established by van den Bergh (1962) and 
Schmidt (1963), and can be described as follows: the number of metal-poor 
stars in the solar vicinity is lower than would be expected on the basis 
of predictions of the simple model (see Tinsley 1980 for a review). Despite 
many theoretical efforts, this problem has remained, initially owing to the lack 
of successful models, and later owing to the multiplicity of them. In the earlier 
attempts to solve this problem, samples were very inhomogeneous, comprising 
stars of different spectral types compiled by several authors. More accurate 
data were obtained years later by Pagel \& Patchett (1975), who established 
a non-biased  sample of 132 G dwarfs within 25 pc of the Sun from the 
nearby star catalogues of Gliese (1969) and Woolley et al. (1970). 

The use of only G dwarfs warrants the homogeneity of the sample, which 
can then be considered as representative of the chemical enrichment history 
of the Galaxy. In fact, G dwarfs are thought to have lifetimes greater than 
or equal to the age of our Galaxy, so that all stars of this type ever 
born in the Galaxy have not yet completed their evolutionary stages. 
Naturally, cooler stars have  even longer lifetimes than G dwarfs. 
However, being less luminous, a sample based on K or M dwarfs would be 
confined to smaller regions, and thus more subject to biases. 

A `G dwarf problem' has not been found in the halo or in the bulge of the 
Galaxy (e.g., Beers, Preston \& Schectman 1985; Laird et al. 1988; Geisler \& 
Friel 1992), and it is not clear whether it exists for low-metallicity globular 
clusters, so that it may be connected with mechanisms operating mainly 
in the disc (Pagel 1987). Therefore, the metallicity distribution of the 
solar neighbourhood is a very important constraint to chemical evolution 
models for our Galaxy. 

The metallicity distributions most often used in the literature are those 
of Pagel (1989) and Sommer-Larsen (1991). Pagel has revised previous data 
of Pagel \& Patchett (1975) by means of a new calibration between the 
ultraviolet excess $\delta (U-B)$ and [Fe/H], and by using the 
oxygen abundance $\log \phi = $ [O/H] as a metallicity index. On the other 
hand, Sommer-Larsen (1991) applied a correction to the distribution of Pagel 
\& Patchett, taking into account the fact that metal-poor stars are generally 
older and have larger scale heights than metal-rich stars. 

The increasing sophistication in chemical evolution models (e.g., 
Sommer-Larsen \& Yoshii 1990; Burkert, Truran \& Hensler 1992; Pardi \& 
Ferrini 1994; Carigi 1994) is supposed to be followed by an improvement of 
the database. Thus, it is desirable to have an up-to-date metallicity 
distribution comprising a larger number of G dwarfs, which could be 
used to discriminate more accurately among the different scenarios 
emerging from these models.

In this paper, we present a new metallicity distribution for the solar 
neighbourhood based on Str\"omgren photometry. In Section 2, we present a 
unbiased sample of G dwarfs from up-to-date photometric and parallax 
catalogues. In Section 3, we apply the corrections introduced by Pagel (1989) 
and Sommer-Larsen (1991) to the metallicity distribution. Some consequences 
for the chemical evolution of the Galaxy and our conclusions follow in 
Section 4.

\section{Sample selection}

\subsection{Selection criteria}

\tx In order to obtain the G dwarf metallicity distribution, we followed the 
criteria of Pagel \& Patchett (1975) for the selection of a non biased 
sample of dwarfs. The stars were selected from the Third Catalogue of Nearby 
Stars (Gliese \& Jahreiss 1991). We have taken all stars with spectral types 
between G0 V and G9 V. For stars with insufficient and/or ambiguous 
classification, we have selected those with visual absolute magnitude 
between 4.5 and $5.13+5.33(B-V)$, to exclude subgiants and white dwarfs, 
respectively (cf. Pagel \& Patchett 1975). A few objects not classified 
as class V stars, but having absolute magnitudes in the interval above, 
were also selected. The total initial sample includes 345 stars. 

We have taken $uvby$ data from the catalogues of Olsen (1993) and Hauck \& 
Mermilliod (1990). We have favoured the first catalogue, when 
a star appeared in both, because the catalogue of Hauck \& Mermilliod lists 
average values from several previous works. 

The number of stars of our initial sample that have photometric data in 
these catalogues is 293. Two of these stars (HD 206827 = HR 8310 and 
HD 218640 = HR 8817) have been excluded from our analysis  in view of  
the following considerations. The first star is classified as G2V in Gliese \& 
Jahreiss (1991) and in the Bright Star Catalogue, but is also referred as an 
F dwarf (cf. Duquennoy \& Mayor 1991). It has a close visual companion (HD 
206826, F6V), and is also suspected of having a spectroscopic companion. 
It has a $b - y$ colour that is too low for a G dwarf ($b - y = 0.241$), although 
it is compatible with that of an early F dwarf. The original paper 
(Oblak 1978) presents $uvby$ colours for HD 206826 B, which Hauck \& 
Mermilliod (1990) probably misassigned to HD 206827. 

The second star, HD 218640, is classified as G2V (Gliese \& Jahreiss 
1991), but its absolute visual magnitude lies far beyond the magnitude 
range we have used to select our sample of G dwarfs. This star also has 
a close visual companion (HD 218641, A2V). The original reference 
(Olsen 1983) gives combined $uvby$ colours for these stars, which were 
assigned by Hauck \& Mermilliod to HD 218640 only.

\subsection{Metallicity calibrations}

\tx The metallicity distributions obtained by Pagel \& Patchett (1975) and 
Pagel (1989) were based on the $UBV$ photometric system, which is 
not very good for the determination of accurate metallicities, since the 
relation between $\delta (U-B)$ and [Fe/H] is not very well defined. 
A metallicity distribution based on Str\"omgren photometry is likely to be 
less subject to errors. Olsen (1984) gives two calibrations in order to 
determine the metallicity:
$${\rm [Fe/H]}_1 = -10\delta m_1 + 0.10 \qquad \eqno\stepeq$$
and 
$${\rm [Fe/H]}_2 = -7.5\delta m_1 + 1.9\delta c_1 -0.04, \qquad 
   \eqno\stepeq$$
where we have introduced
$$\delta m_1 = m_1^{\rm Hyades} - m_1^{\rm star}\eqno\stepeq$$
and
$$\delta c_1 = c_1^{\rm Hyades} - c_1^{\rm star}.\eqno\stepeq$$
Here $m_1$ and $c_1$ are the standard $uvby$ indices. 
Equations (1) and (2) will be called first and second calibrations, respectively. 
The Hyades standard curves $m_1\times(b-y)$ and $c_1\times(b-y)$ for G  
dwarfs were taken from Crawford (1975) and Olsen (1984). The adopted 
sample is shown in Table 1, where we list the HD number (column 1), 
the spectral type (column 2), the photometric indices $(b-y)$ (column 3), 
$m_1$ (column 4), $c_1$ (column 5), $\delta m_1$  (column 6), and 
$\delta c_1$ (column 7). The metallicities as given by equations (1) and 
(2) are given in columns 8 and 9, respectively. The average errors derived 
from the metallicity calibration are in the range $0.15 - 0.17$ dex 
(cf. Olsen 1984). 

A more recent metallicity calibration has been presented by Schuster 
\& Nissen (1989), which makes use of the standard $uvby$ indices 
themselves:
\bigskip\noindent
$${\rm [Fe/H]}_3 = 1.052 - 73.21 m_1 + 280.9 m_1(b-y) \ + $$
\smallskip
$$\phantom{{\rm [Fe/H]}_3 =} 333.95 m_1^2(b-y)- 595.5 m_1(b-y)^2 \ + $$
\smallskip
$$\phantom{{\rm [Fe/H]}_3 =} [5.486 - 41.62 m_1 - 7.963 (b-y)]\log (m_1-c_3),
\eqno\startsubeq$$
\bigskip
\noindent 
for $0.22 \leq (b-y) < 0.375$, and
\bigskip\noindent
$${\rm [Fe/H]}_3 = -2.0965 + 22.45 m_1 - 53.8 m_1^2 -$$
\smallskip
$$\phantom{{\rm [Fe/H]}_3 =} 62.04 m_1 (b-y) + 145.5 m_1^2 (b-y) \ + $$
\smallskip
$$\phantom{{\rm [Fe/H]}_3 =}[85.1 m_1 - 13.8 c_1 - 137.2 m_1^2] c_1,
\eqno\stepsubeq$$
\bigskip
\noindent
for $ 0.375 \leq (b-y) \leq 0.59$, which we shall call calibration 3. In 
equation (5a) we have introduced
$$c_3 = 0.6322 - 3.58 (b-y) + 5.20 (b-y)^2. \eqno\stepeq$$
The average uncertainty for this calibration is $\simeq 0.16$ dex, and the 
metallicities derived are shown in column 10 of Table~1. 

A comparison of [Fe/H]$_1$ and [Fe/H]$_3$ shows that calibration 1 yields 
smaller metallicities for the redder stars ($b-y \geq 0.45$) than 
calibration 3. This discrepancy may arise from the fact that equation (1) 
assumes a simple linear dependence of the metallicity on $\delta m_1$, 
which may be a poor approximation at lower metallicities (cf. Olsen 1984). 
\beginfigure*{1}
\vskip 19 cm
\caption{{\bf Figure 1.} Photometric versus spectroscopic 
    abundances for 79 stars of our sample. (a) Calibration 1 (equation 1); 
    (b) calibration 2 (equation 2); (c) calibration 3 (equations 5).}
\endfigure

In Fig. 1, we compare the derived photometric abundances with spectroscopic 
abundances available in the literature for 79 stars of our sample. The 
spectroscopic metallicities [Fe/H]$_{\rm S}$ were taken from Edvardsson et al. 
(1993) and Cayrel de Strobel et al. (1992), and are shown in column 11 of 
Table~1. We have favoured the data from Edvardsson et al. for stars appearing 
in both works, as the values from Cayrel de Strobel et al. have been compiled 
from several different sources. Figs. 1(a), 1(b) and 1(c) show the results 
of calibrations 1, 2  and 3, respectively. 
The dispersion is very large, but may be partially due to inhomogeneities 
of the catalogue of Cayrel de Strobel et al. (1992). It can be seen that 
calibrations~1 and 3  yield abundances in better agreement with the 
spectroscopic abundances than calibration~2. The agreement is better for 
calibration~3, so that we will adopt the photometric abundances as given 
by equations~(5). 
\begintable*{1}
\caption{{\bf Table 1.} $uvby$ indices and abundances for a 
sample of G dwarfs.}
{\halign{%
\hfil \rm #  & \quad \rm \hfil # \hfil & \quad $#$ \hfil 
& \quad $#$ \hfil & \quad $#$ \hfil & \quad \hfil $#$
 & \quad \hfil$#$ & \quad \hfil$#$ & \quad \hfil$#$& \quad \hfil$#$ & 
\quad \hfil$#$ \cr
HD \  & MK type & b-y & \ \ m_1 & \ \ c_1 & \delta m_1 & \delta c_1\  & 
{\rm [Fe/H]}_1 & {\rm [Fe/H]}_2 & {\rm [Fe/H]}_3 & {\rm [Fe/H]}_{\rm S} \cr
\noalign{\vskip 10 pt}
1237 & G6 V  & 0.459  & 0.289  & 0.300 & 0.011  & -0.061  & -0.01  & -0.24 
& -0.06 & \cr
1273 & G2 V & 0.410 & 0.184 & 0.250 & 0.046 & 0.019 & -0.36  & -0.35 
& -0.56& \cr 
1388 & G2 V & 0.379 & 0.191 & 0.345 & 0.014 & -0.036 & -0.04 & -0.21 
& -0.01 & \cr 
1461 & G5 IV-V & 0.421 & 0.244 & 0.360 & -0.001 & -0.102 & 0.11 & -0.23 
&  0.20 & 0.41 \cr 
1835 & G2 V & 0.411 & 0.225 & 0.353 & 0.006 & -0.085 & 0.04 & -0.25 
&  0.11 & 0.20 \cr 
3405 & G3 V & 0.397 & 0.204 & 0.362 & 0.014 & -0.077 & -0.04 & -0.29 
&  0.03 & \cr 
3443 & G7 V & 0.435 & 0.252 & 0.290 & 0.009 & -0.043 & 0.01 & -0.19 
& -0.11 & -0.16 \cr 
3795 & G3 V & 0.443 & 0.217 & 0.282 & 0.056 & -0.039 & -0.46 & -0.53 
& -0.40 & \cr 
3823 & G1 V & 0.364 & 0.150 & 0.349 & 0.047 & -0.020 & -0.37 & -0.43 
& -0.48 & \cr 
4208 & G5 V & 0.404 & 0.227 & 0.278 & -0.003 & -0.002 & 0.13 & 0.15 
& -0.09 & \cr 
4307 & G0 V & 0.387 & 0.173 & 0.348 & 0.037 & -0.050 & -0.27 & -0.41 
& -0.27 & -0.28 \cr 
4308 & G3 V & 0.409 & 0.187 & 0.302 & 0.042 & -0.032 & -0.32 & -0.42 
& -0.35 & -0.40 \cr 
4614 & G3 V & 0.372 & 0.185 & 0.275 & 0.016 & 0.043 & -0.06 & 0.02 
& -0.10 & -0.31 \cr 
4747 & G9 V & 0.460 & 0.295 & 0.275 & 0.007 & -0.037 & 0.03 & -0.16 
& -0.17 & \cr 
5817 & dG2 & 0.398 & 0.156 & 0.254 & 0.063 & 0.029 & -0.53 & -0.46 
& -0.71 & \cr 
6582 & G5 VI & 0.434 & 0.199 & 0.215 & 0.061 & 0.033 & -0.51 & -0.43 
& -0.78 & -0.90\cr 
6880 & G8/K0 V & 0.318 & 0.154 & 0.480 & 0.026 & -0.096 & -0.16 & -0.42 
& -0.21 & \cr 
9540 & G8 V & 0.451 & 0.291 & 0.294 & -0.005 & -0.054 & 0.15 & -0.10 
& -0.03 & \cr
10145 & G5 V & 0.421 & 0.232 & 0.326 & 0.011 & -0.068 & -0.01 & -0.25 
&  0.02 & \cr 
10700 & G8 Vp & 0.435 & 0.263 & 0.238 & -0.002 & 0.009 & 0.12 & 0.13 
& -0.35 & -0.58 \cr 
10800 & G2 V & 0.392 & 0.195 & 0.303 & 0.019 & -0.012 & -0.09 & -0.20 
& -0.16 & \cr 
11112 & G4 V & 0.409 & 0.200 & 0.400 & 0.029 & -0.130 & -0.19 & -0.51
& -0.05 & \cr 
11131 & dG1 & 0.394 & 0.206 & 0.292 & 0.010 & -0.003 & 0.00 & -0.12 
& -0.12 & -0.06 \cr 
12051 & dG7 & 0.475 & 0.309 & 0.372 & 0.021 & -0.134 & -0.11 & -0.45 
&  0.20 & \cr 
12759 & G3 V & 0.433 & 0.227 & 0.333 & 0.031 & -0.085 & -0.21 & -0.44 
& -0.08 & \cr 
13043 & G2 V & 0.393 & 0.196 & 0.372 & 0.019 & -0.082 & -0.09 & -0.34 
& -0.01 & \cr 
13974 & G0 Ve & 0.386 & 0.191 & 0.254 & 0.018 & 0.045 & -0.08 & 0.01 
& -0.33 & -0.30 \cr 
14412 & G5 V & 0.438 & 0.258 & 0.229 & 0.008 & 0.016 & 0.02 & 0.09 
& -0.44 & \cr 
14802 & G1 V & 0.385 & 0.188 & 0.373 & 0.021 & -0.072 & -0.11 & -0.33 
& -0.05 & \cr 
15335 & G0 V & 0.387 & 0.160 & 0.370 & 0.050 & -0.072 & -0.40 & -0.55 
& -0.43 & -0.22 \cr 
16397 & G1 V & 0.387 & 0.159 & 0.278 & 0.051 & 0.020 & -0.41 & -0.39 
& -0.55 & \cr 
16623 & G2 V-VI & 0.372 & 0.152 & 0.372 & 0.049 & -0.054 & -0.39 & -0.51 
& -0.50 & \cr 
16784 & G0 V & 0.378 & 0.142 & 0.293 & 0.062 & 0.017 & -0.52 & -0.47 
& -0.67 & \cr 
17169 & G5 V & 0.468 & 0.220 & 0.339 & 0.096 & -0.101 & -0.86 & -0.96 
& -0.34 & \cr 
18757 & G4 V & 0.413 & 0.199 & 0.303 & 0.035 & -0.037 & -0.25 & -0.37 
& -0.26 & \cr
18803 & G8 V & 0.425 & 0.270 & 0.328 & -0.022 & -0.074 & 0.32 & -0.01 
&  0.20 & \cr 
19034 & dG5 & 0.416 & 0.214 & 0.263 & 0.023 & -0.001 & -0.13 & -0.21 
& -0.32 & \cr 
19467 & G5 V & 0.409 & 0.200 & 0.337 & 0.029 & -0.067 & -0.19 & -0.39 
& -0.13 & \cr 
20407 & G3 V & 0.373 & 0.162 & 0.280 & 0.039 & 0.037 & -0.29 & -0.26 
& -0.40 & \cr 
20619 & G1.5 V & 0.405 & 0.214 & 0.269 & 0.011 & 0.006 & -0.01 & -0.11 
& -0.22 & -0.45 \cr 
20630 & G5 Ve & 0.415 & 0.242 & 0.310 & -0.006 & -0.047 & 0.16 & 0.24 
&  0.06 & 0.04 \cr 
20727 & dG2 & 0.436 & 0.185 & 0.358 & 0.078 & -0.112 & -0.68 & -0.83 
& -0.43 & \cr 
20766 & G2 V & 0.404 & 0.196 & 0.289 & 0.028 & -0.013 & -0.18 & -0.28 
& -0.27 & -0.35 \cr 
20794 & G5 V & 0.439 & 0.235 & 0.285 & 0.032 & -0.040 & -0.22 & -0.36 
& -0.25 & -0.54 \cr 
20807 & G2 V & 0.383 & 0.177 & 0.297 & 0.030 & 0.006 & -0.20 & -0.26 
& -0.29 & -0.23 \cr 
21019 & G2 V & 0.449 & 0.203 & 0.298 & 0.080 & -0.057 & -0.70 & -0.75 
& -0.48 & \cr 
21411 & G8 V & 0.440 & 0.253 & 0.258 & 0.015 & -0.014 & -0.05 & -0.18 
& -0.30 & \cr 
24293 & G5 V & 0.412 & 0.211 & 0.332 & 0.022 & -0.065 & -0.12 & -0.33 
& -0.07 & \cr 
24365 & G8 V & 0.506 & 0.294 & 0.303 & 0.104 & -0.062 & -0.94 & -0.94 
& -0.29 & \cr 
24892 & G8 V & 0.461 & 0.240 & 0.313 & 0.063 & -0.075 & -0.53 & -0.66 
& -0.24 & \cr
25680 & G5 V & 0.401 & 0.200 & 0.331 & 0.022 & -0.051 & -0.12 & -0.30 
& -0.10 & \cr 
26491 & G3 V & 0.406 & 0.188 & 0.329 & 0.038 & -0.056 & -0.28 & -0.43 
& -0.25 & -0.18 \cr
28255 & G4 V & 0.405 & 0.217 & 0.332 & 0.008 & -0.057 & 0.02 & -0.21 
&  0.03 & \cr 
29231 & G8 V & 0.469 & 0.307 & 0.327 & 0.011 & -0.089 & -0.01 & -0.29 
&  0.06 & \cr 
30003 & G5 V & 0.427 & 0.230 & 0.339 & 0.020 & -0.086 & -0.10 & -0.36 
&  0.00 & \cr 
30495 & G1 V & 0.395 & 0.207 & 0.326 & 0.009 & -0.039 & 0.01 & -0.18 
& -0.01 & 0.10 \cr 
30649 & G1 IV-V & 0.384 & 0.155 & 0.293 & 0.053 & 0.009 & -0.43 & -0.42 
& -0.55 & -0.51 \cr 
32387 & G8 V & 0.501 & 0.247 & 0.291 & 0.140 & -0.050 & -1.30 & -1.18 
& -0.55 & \cr 
32778 & G5 V & 0.395 & 0.180 & 0.261 & 0.036 & 0.026 & -0.26 & -0.26 
& -0.45 & \cr 
32923 & G4 V & 0.415 & 0.197 & 0.332 & 0.039 & -0.069 & -0.29 & -0.46 
& -0.21 & -0.20 \cr 
}} 
\endtable
\begintable*{2}
\caption{{\bf Table 1} -- {\it continued}}
{\halign{%
\hfil \rm #  & \quad \rm \hfil # \hfil & \quad $#$ \hfil 
& \quad $#$ \hfil & \quad $#$ \hfil & \quad \hfil $#$
 & \quad \hfil$#$ & \quad \hfil$#$ & \quad \hfil$#$ & \quad \hfil$#$
& \quad \hfil$#$ \cr
HD \  & MK type & b-y & \ \ m_1 & \ \ c_1 & \delta m_1 & \delta c_1\  & 
{\rm [Fe/H]}_1 & {\rm [Fe/H]}_2 & {\rm [Fe/H]}_3& {\rm [Fe/H]}_{\rm S} \cr
\noalign{\vskip 10 pt}
33811 & G8 IV-V & 0.465 & 0.283 & 0.409 & 0.028 & -0.171 & -0.18 & -0.57 
&  0.25 & \cr 
34101 & G8 V & 0.444 & 0.249 & 0.329 & 0.026 & -0.087 & -0.16 & -0.40 
& -0.02 & \cr 
34721 & G0 V & 0.361 & 0.174 & 0.364 & 0.021 & -0.032 & -0.11 & -0.26 
& -0.17 & \cr 
35956 & G0 V & 0.377 & 0.182 & 0.321 & 0.022 & -0.009 & -0.12 & -0.22 
& -0.14 & \cr 
36435 & G5 Ve & 0.454 & 0.300 & 0.263 & -0.009 & -0.024 & 0.19 & 0.21 
& -0.20 & \cr 
37655 & G0 V & 0.380 & 0.176 & 0.373 & 0.029 & -0.066 & -0.19 & -0.39 
& -0.17 & \cr 
37706 & G5 V & 0.454 & 0.297 & 0.240 & -0.006 & -0.001 & 0.16 & 0.00 
& -0.35 & \cr 
38858 & G2 V & 0.402 & 0.192 & 0.288 & 0.031 & -0.010 & -0.21 & -0.29 
& -0.30 & \cr 
39091 & G1 V & 0.370 & 0.190 & 0.367 & 0.010 & -0.046 & 0.00 & -0.20 
& -0.02 & 0.00 \cr
39587 & G0 V & 0.378 & 0.194 & 0.307 & 0.010 & 0.003 & 0.00 & -0.11 
& -0.06 & -0.03 \cr 
41330 & G0 V & 0.389 & 0.161 & 0.348 & 0.051 & -0.053 & -0.41 & -0.52 
& -0.44 & -0.24 \cr 
42250 & dG7 & 0.469 & 0.287 & 0.333 & 0.031 & -0.095 & -0.21 & -0.46 
&  0.02 & \cr 
42618 & G4 V & 0.403 & 0.216 & 0.303 & 0.007 & -0.026 & 0.03 & -0.15 
& -0.06 & \cr 
42807 & G2 V & 0.418 & 0.226 & 0.289 & 0.013 & -0.029 & -0.03 & -0.19 
& -0.14 & \cr 
43162 & G5 V & 0.428 & 0.246 & 0.303 & 0.006 & -0.051 & 0.04 & -0.18 
& -0.03 & \cr 
43834 & G5 V & 0.443 & 0.262 & 0.340 & 0.011 & -0.097 & -0.01 & -0.31 
&  0.10 & 0.01 \cr 
44120 & G3 V & 0.378 & 0.174 & 0.408 & 0.030 & -0.098 & -0.20 & -0.45 
& -0.18 & \cr 
45184 & G0 V & 0.394 & 0.207 & 0.338 & 0.009 & -0.049 & 0.01 & -0.20 
&  0.03 & \cr 
48189 & G0 V & 0.392 & 0.195 & 0.309 & 0.019 & -0.018 & -0.09 & -0.22 
& -0.14 & \cr 
48938 & G2 V & 0.354 & 0.139 & 0.333 & 0.053 & 0.008 & -0.43 & -0.42 
& -0.55 & -0.38 \cr 
50692 & G0 V & 0.382 & 0.167 & 0.329 & 0.040 & -0.024 & -0.30 & -0.38 
& -0.33 & \cr 
51608 & G7 V & 0.461 & 0.297 & 0.325 & 0.006 & -0.087 & 0.04 & -0.25 
&  0.07 & \cr 
52711 & G4 V & 0.388 & 0.172 & 0.333 & 0.039 & -0.036 & -0.29 & -0.40 
& -0.30 & -0.15 \cr 
53705 & G3 V & 0.396 & 0.173 & 0.331 & 0.044 & -0.045 & -0.34 & -0.46 
& -0.35 & \cr 
55575 & G0 V & 0.384 & 0.150 & 0.307 & 0.058 & -0.005 & -0.48 & -0.49 
& -0.59 & -0.28 \cr 
56274 & G2 V & 0.396 & 0.148 & 0.266 & 0.069 & 0.020 & -0.59 & -0.52 
& -0.75 & -0.70 \cr
59468 & G5 IV-V & 0.433 & 0.253 & 0.323 & 0.005 & -0.075 & 0.05 & -0.22 
&  0.05 & \cr 
59967 & G4 V & 0.399 & 0.200 & 0.301 & 0.020 & -0.019 & -0.10 & -0.22 
& -0.17 & \cr 
61994 & dG5 & 0.445 & 0.247 & 0.298 & 0.029 & -0.056 & -0.19 & -0.37 
& -0.16 & \cr 
62613 & G8 V & 0.450 & 0.261 & 0.293 & 0.023 & -0.053 & -0.13 & -0.31 
& -0.14 & \cr 
63077 & G0 V & 0.371 & 0.137 & 0.278 & 0.063 & 0.042 & -0.53 & -0.43 
& -0.65 & -0.78 \cr 
64090 & G2 VI & 0.436 & 0.104 & 0.106 & 0.159 & 0.140 & -1.49 & -0.96 
& -1.84 & -1.92 \cr 
64184 & G5 V & 0.425 & 0.210 & 0.305 & 0.038 & -0.051 & -0.28 & -0.42 
& -0.24 & \cr 
64606 & G8 V & 0.454 & 0.211 & 0.208 & 0.080 & 0.031 & -0.70 & -0.58 
& -0.89 & -0.99 \cr 
65583 & G8 V & 0.455 & 0.220 & 0.229 & 0.073 & 0.010 & -0.63 & -0.57 
& -0.72 & -0.60 \cr 
65721 & G6 V & 0.453 & 0.248 & 0.268 & 0.041 & -0.028 & -0.31 & -0.40 
& -0.35 & \cr 
65907 & G2 V & 0.370 & 0.153 & 0.324 & 0.047 & -0.003 & -0.37 & -0.40 
& -0.48 & \cr 
67458 & G4 IV-V & 0.375 & 0.189 & 0.302 & 0.013 & 0.012 & -0.03 & -0.12 
& -0.10 & \cr 
68017 & G4 V & 0.420 & 0.194 & 0.264 & 0.048 & -0.005 & -0.38 & -0.41 
& -0.49 & \cr 
69565 & G8/K0 V & 0.554 & 0.299 & 0.421 & 0.218 & -0.197 & -2.08 & -2.05 
& -0.16 & \cr
69830 & G7.5 V & 0.457 & 0.297 & 0.314 & -0.001 & -0.075 & 0.11 & -0.18 
&  0.05 & \cr 
70642 & G6 V & 0.435 & 0.252 & 0.350 & 0.009 & -0.103 & 0.01 & -0.30 
&  0.13 & \cr 
71148 & G5 V & 0.402 & 0.193 & 0.348 & 0.030 & -0.070 & -0.20 & -0.39 
& -0.14 & \cr
71334 & G4 V & 0.415 & 0.210 & 0.324 & 0.026 & -0.061 & -0.16 & -0.35 
& -0.12 & \cr 
72905 & G1 V & 0.403 & 0.188 & 0.304 & 0.035 & -0.027 & -0.25 & -0.36 
& -0.29 & -0.08 \cr 
73524 & G4 IV-V & 0.385 & 0.183 & 0.385 & 0.026 & -0.084 & -0.16 & -0.39 
& -0.10 & \cr 
73752 & G3 V & 0.447 & 0.277 & 0.394 & 0.002 & -0.153 & 0.08 & -0.35 
& 0.30 & \cr 
74842 & G5 V & 0.454 & 0.269 & 0.268 & 0.022 & -0.029 & -0.12 & -0.26 
& -0.26 & \cr 
75732 & G8 V & 0.536 & 0.357 & 0.415 & 0.115 & -0.180 & -1.05 & -1.24 
&  0.10 & 0.30 \cr 
76151 & G3 V & 0.412 & 0.234 & 0.336 & -0.001 & -0.069 & 0.11 & -0.16 
&  0.12 & 0.01 \cr 
77137 & G5 V & 0.437 & 0.233 & 0.365 & 0.031 & -0.119 & -0.21 & -0.50 
&  0.03 & \cr 
78643 & G1 V & 0.368 & 0.166 & 0.385 & 0.033 & -0.061 & -0.23 & -0.40 
& -0.32 & \cr 
81809 & G2 V & 0.416 & 0.182 & 0.344 & 0.055 & -0.082 & -0.45 & -0.61 
& -0.35 & -0.31 \cr 
82885 & G8 V & 0.473 & 0.304 & 0.372 & 0.022 & -0.134 & -0.12 & -0.46 
&  0.19 & 0.00 \cr 
86728 & G2 Va & 0.416 & 0.234 & 0.388 & 0.003 & -0.126 & 0.07 & -0.30 
&  0.22 & 0.10 \cr 
88261 & G3 V-VI & 0.389 & 0.150 & 0.286 & 0.062 & 0.009 & -0.52 & -0.49 
& -0.65 & \cr 
88725 & G1 V & 0.404 & 0.147 & 0.264 & 0.077 & 0.012 & -0.67 & -0.60 
& -0.82 & \cr 
88742 & G1 V & 0.386 & 0.174 & 0.332 & 0.035 & -0.033 & -0.25 & -0.37 
& -0.27 & \cr 
89269 & G5 V & 0.420 & 0.208 & 0.292 & 0.034 & -0.033 & -0.24 & -0.36 
& -0.27 & \cr 
89906 & dG2 & 0.422 & 0.208 & 0.253 & 0.036 & 0.004 & -0.26 & -0.30 
& -0.45 & \cr
 }}
\endtable
\begintable*{3}
\caption{{\bf Table 1} -- {\it continued}}
{\halign{%
\hfil \rm #  & \quad \rm \hfil # \hfil & \quad $#$ \hfil 
& \quad $#$ \hfil & \quad $#$ \hfil & \quad \hfil $#$
 & \quad \hfil$#$ & \quad \hfil$#$ & \quad \hfil$#$ & \quad \hfil$#$
& \quad \hfil$#$ \cr
HD \ & MK type & b-y & \ \ m_1 & \ \ c_1 & \delta m_1 & \delta c_1\  & 
{\rm [Fe/H]}_1 & {\rm [Fe/H]}_2 & {\rm [Fe/H]}_3& {\rm [Fe/H]}_{\rm S} \cr
\noalign{\vskip 10 pt}
90156 & G5 V & 0.424 & 0.208 & 0.296 & 0.038 & -0.041 & -0.28 & -0.41 
& -0.28 & \cr 
94340 & G3/5 V & 0.406 & 0.192 & 0.384 & 0.034 & -0.111 & -0.24 & -0.51 
& -0.13 & \cr 
94518 & G2 V & 0.382 & 0.174 & 0.277 & 0.033 & 0.028 & -0.23 & -0.23 
& -0.37 & \cr 
95128 & G0 V & 0.392 & 0.203 & 0.337 & 0.011 & -0.046 & -0.01 & -0.21 
&  0.01 & 0.01 \cr 
96700 & G2 V & 0.398 & 0.163 & 0.325 & 0.056 & -0.042 & -0.46 & -0.54 
& -0.49 & \cr 
97334 & G0 V & 0.392 & 0.210 & 0.311 & 0.004 & -0.020 & 0.06 & -0.11 
& -0.01 & \cr
97343 & G8/K0 V & 0.462 & 0.280 & 0.345 & 0.025 & -0.107 & -0.15 & -0.43 
&  0.08 & \cr 
98230 & G0 Ve & 0.377 & 0.180 & 0.293 & 0.024 & 0.019 & -0.14 & -0.18 
& -0.23 & -0.12 \cr 
98281 & G8 V & 0.457 & 0.254 & 0.288 & 0.042 & -0.049 & -0.32 & -0.45 
& -0.25 & \cr 
100180 & G0 V & 0.373 & 0.174 & 0.353 & 0.027 & -0.036 & -0.17 & -0.31 
& -0.26 &  \cr 
101177 & G0 V & 0.371 & 0.182 & 0.310 & 0.018 & 0.010 & -0.08 & -0.16 
& -0.14 & \cr 
101259 & G6/8 V & 0.508 & 0.229 & 0.315 & 0.174 & -0.074 & -1.64 & -1.49 
& -0.61 & \cr 
101501 & G8 Ve & 0.445 & 0.264 & 0.290 & 0.012 & -0.048 & -0.02 & -0.22 
& -0.11 & 0.03 \cr 
101563 & G0 V & 0.416 & 0.196 & 0.342 & 0.041 & -0.080 & -0.31 & -0.50 
& -0.21 & \cr 
102365 & G5 V & 0.418 & 0.199 & 0.278 & 0.040 & -0.018 & -0.30 & -0.38 
& -0.38 & -0.70 \cr 
102438 & G5 V & 0.433 & 0.210 & 0.281 & 0.048 & -0.033 & -0.38 & -0.46 
& -0.39 & \cr 
102540 & G5/6 V & 0.469 & 0.227 & 0.312 & 0.091 & -0.074 & -0.81 & -0.87 
& -0.38 & \cr
103095 & G8 VI & 0.482 & 0.224 & 0.166 & 0.121 & 0.073 & -1.11 & -0.81 
& -1.31 & -1.40 \cr 
103431 & dG7 & 0.439 & 0.249 & 0.287 & 0.018 & -0.042 & -0.08 & -0.26 
& -0.16 & \cr 
103432 & dG6 & 0.420 & 0.224 & 0.287 & 0.018 & -0.028 & -0.08 & -0.23 
& -0.18 & \cr 
103493 & G5 V & 0.401 & 0.194 & 0.288 & 0.028 & -0.008 & -0.18 & -0.26 
& -0.27 & \cr 
104471 & G0 V & 0.370 & 0.191 & 0.363 & 0.009 & -0.042 & 0.01 & -0.18 
&  0.00 & \cr 
105590 & G2 V & 0.420 & 0.220 & 0.320 & 0.022 & -0.061 & -0.12 & -0.32 
& -0.08 & -0.04 \cr 
106116 & G4 V & 0.431 & 0.249 & 0.359 & 0.007 & -0.109 & 0.03 & -0.30 
&  0.16 & \cr 
106156 & G8 V & 0.471 & 0.329 & 0.341 & -0.007 & -0.103 & 0.17 & -0.19 
&  0.15 & \cr 
108754 & G8 V & 0.435 & 0.217 & 0.237 & 0.044 & 0.010 & -0.34 & -0.35 
& -0.56 & \cr 
108799 & G0 V & 0.381 & 0.174 & 0.319 & 0.032 & -0.013 & -0.22 & -0.31 
& -0.26 & \cr 
109358 & G0 V & 0.385 & 0.182 & 0.296 & 0.027 & 0.005 & -0.17 & -0.23 
& -0.25 & -0.19 \cr 
110010 & dG2 & 0.395 & 0.228 & 0.387 & -0.012 & -0.100 & 0.22 & -0.14 
&  0.31 & \cr 
110897 & G0 V & 0.374 & 0.149 & 0.284 & 0.053 & 0.032 & -0.43 & -0.38 
& -0.55 & -0.59 
\cr
111031 & G5 V & 0.426 & 0.250 & 0.376 & -0.001 & -0.123 & 0.11 & -0.27 
&  0.24 & \cr 
111395 & G7 V & 0.438 & 0.241 & 0.334 & 0.025 & -0.089 & -0.15 & -0.39 
& -0.01 & \cr 
111515 & G8 V & 0.437 & 0.201 & 0.241 & 0.063 & 0.005 & -0.53 & -0.50 
& -0.65 & \cr 
112164 & G1 V & 0.402 & 0.197 & 0.460 & 0.026 & -0.182 & -0.16 & -0.58 
& -0.06 & 0.24 \cr
114260 & G6 V & 0.452 & 0.248 & 0.310 & 0.040 & -0.070 & -0.30 & -0.47 
& -0.15 & \cr 
114613 & G3 V & 0.441 & 0.235 & 0.390 & 0.035 & -0.146 & -0.25 & -0.58 
&  0.07 & \cr 
114710 & G0 V & 0.370 & 0.191 & 0.337 & 0.009 & -0.016 & 0.01 & -0.13 
&  0.00 & 0.03 \cr 
115383 & G0 V & 0.376 & 0.191 & 0.383 & 0.012 & -0.070 & -0.02 & -0.26 
&  0.05 & 0.10 \cr 
115617 & G6 V & 0.433 & 0.256 & 0.328 & 0.002 & -0.080 & 0.08 & -0.21 
&  0.09 & -0.03 \cr 
116442 & G5 V & 0.467 & 0.310 & 0.251 & 0.005 & -0.013 & 0.05 & 0.15 
& -0.32 & \cr 
116443 & G5 V & 0.493 & 0.369 & 0.260 & 0.000 & -0.020 & 0.10 & 0.18 
& -0.28 & \cr 
117043 & dG6 & 0.458 & 0.272 & 0.352 & 0.026 & -0.113 & -0.16 & -0.45 
&  0.10 & \cr 
117176 & G2.5 Va & 0.451 & 0.221 & 0.354 & 0.065 & -0.114 & -0.55 & -0.74 
& -0.19 & -0.11 \cr 
117635 & G9 V & 0.474 & 0.284 & 0.249 & 0.044 & -0.011 & -0.34 & -0.39 
& -0.44 & \cr 
117939 & G4 V & 0.424 & 0.191 & 0.303 & 0.055 & -0.048 & -0.45 & -0.55 
& -0.40 & \cr 
120690 & G5 V & 0.433 & 0.236 & 0.315 & 0.022 & -0.067 & -0.12 & -0.33 
& -0.08 & \cr 
121384 & G8 V & 0.482 & 0.234 & 0.302 & 0.111 & -0.063 & -1.01 & -0.99 
& -0.46 & \cr 
121849 & G5 V & 0.432 & 0.202 & 0.294 & 0.055 & -0.045 & -0.45 & -0.54 
& -0.39 & \cr 
122742 & G8 V & 0.452 & 0.264 & 0.317 & 0.024 & -0.077 & -0.14 & -0.36 
& -0.03 & \cr 
122862 & G1 V & 0.376 & 0.163 & 0.371 & 0.040 & 0.058 & -0.30 & -0.45 
& -0.32 & \cr 
123505 & G9 V & 0.475 & 0.255 & 0.234 & 0.075 & 0.004 & -0.65 & -0.60 
& -0.66 & \cr
124580 & G4 V & 0.384 & 0.177 & 0.314 & 0.031 & -0.012 & -0.21 & -0.30 
& -0.26 & \cr 
125184 & G8 V & 0.454 & 0.247 & 0.413 & 0.044 & -0.174 & -0.34 & -0.70 
&  0.11 & 0.13 \cr 
126053 & G1 V & 0.406 & 0.183 & 0.287 & 0.043 & -0.014 & -0.33 & -0.39 
& -0.41 & \cr 
126525 & G5 V & 0.426 & 0.222 & 0.310 & 0.027 & -0.057 & -0.17 & -0.35 
& -0.14 & \cr 
127334 & G5 V & 0.441 & 0.245 & 0.366 & 0.025 & -0.122 & -0.15 & -0.46 
&  0.09 & 0.05 \cr 
127356 & G5 V & 0.448 & 0.217 & 0.261 & 0.064 & -0.020 & -0.54 & -0.56 
& -0.53 & \cr 
128400 & G5 V & 0.438 & 0.243 & 0.306 & 0.023 & -0.061 & -0.13 & -0.32 
& -0.10 & \cr 
128620 & G2 V & 0.438 & 0.248 & 0.373 & 0.018 & -0.128 & -0.08 & -0.41 
&  0.15 & 0.15 \cr 
129333 & dG0 & 0.408 & 0.202 & 0.301 & 0.026 & -0.030 & -0.16 & -0.29 
& -0.21 & \cr  }}
\endtable
\begintable*{4}
\caption{{\bf Table 1} -- {\it continued}}
{\halign{%
\hfil \rm #  & \quad \rm \hfil # \hfil & \quad $#$ \hfil 
& \quad $#$ \hfil & \quad $#$ \hfil & \quad \hfil $#$
 & \quad \hfil$#$ & \quad \hfil$#$ & \quad \hfil$#$ & \quad \hfil$#$
& \quad \hfil$#$ \cr
HD \ & MK type & b-y & \ \ m_1 & \ \ c_1 & \delta m_1 & \delta c_1\  & 
{\rm [Fe/H]}_1 & {\rm [Fe/H]}_2 & {\rm [Fe/H]}_3& {\rm [Fe/H]}_{\rm S} \cr
\noalign{\vskip 10 pt}
130307 & G8 V & 0.521 & 0.405 & 0.275 & 0.030 & -0.035 & -0.20 & -0.33 
& -0.24 & \cr 
130948 & G2 V & 0.383 & 0.179 & 0.326 & 0.028 & -0.023 & -0.18 & -0.30 
& -0.20 & 0.20 \cr 
131923 & G5 V & 0.443 & 0.228 & 0.361 & 0.045 & -0.118 & -0.35 & -0.60 
& -0.06 & \cr 
134319 & dG5 & 0.419 & 0.222 & 0.276 & 0.018 & -0.017 & -0.08 & -0.21 
& -0.23 & \cr 
134987 & G5 V & 0.434 & 0.256 & 0.375 & 0.004 & -0.127 & 0.06 & -0.31 
&  0.23 & \cr 
135101 & G5 V & 0.436 & 0.217 & 0.369 & 0.046 & -0.123 & -0.36 & -0.62 
& -0.09 & \cr 
135101 & G7 V & 0.460 & 0.251 & 0.353 & 0.051 & -0.115 & -0.41 & -0.64 
& -0.03 & \cr 
136352 & G2 V & 0.409 & 0.181 & 0.298 & 0.048 & -0.028 & -0.38 & -0.46 
& -0.41 & -0.40 \cr
137107 & G2 V & 0.369 & 0.185 & 0.341 & 0.014 & -0.019 & -0.04 & -0.18 
& -0.08 & \cr 
137392 & G1 V & 0.383 & 0.196 & 0.370 & 0.011 & -0.067 & -0.01 & -0.25 
&  0.05 & 0.05 \cr 
137676 & G5 V & 0.476 & 0.226 & 0.293 & 0.106 & -0.055 & -0.96 & -0.94 
& -0.51 & \cr 
140538 & G5 V & 0.425 & 0.228 & 0.336 & 0.020 & -0.082 & -0.10 & -0.34 
& -0.01 & \cr 
140901 & G6 V & 0.436 & 0.268 & 0.324 & -0.005 & -0.078 & 0.15 & -0.15 
&  0.11 & \cr 
141004 & G0 V & 0.385 & 0.199 & 0.354 & 0.010 & -0.053 & 0.00 & -0.21 
&  0.05 & -0.04 \cr 
141272 & G8 V & 0.479 & 0.324 & 0.304 & 0.014 & -0.065 & -0.04 & -0.27 
& -0.05 & \cr 
142267 & G1 V & 0.393 & 0.159 & 0.276 & 0.056 & 0.014 & -0.46 & -0.43 
& -0.59 & -0.28 \cr 
143761 & G2 V & 0.394 & 0.183 & 0.322 & 0.033 & -0.033 & -0.23 & -0.35 
& -0.24 & -0.26 \cr 
144009 & G8 V & 0.450 & 0.251 & 0.327 & 0.033 & -0.087 & -0.23 & -0.45 
& -0.05 & \cr 
144087 & G8 V & 0.450 & 0.265 & 0.338 & 0.019 & -0.098 & -0.09 & -0.37 
&  0.06 & \cr 
144179 & G9 V & 0.481 & 0.308 & 0.254 & 0.035 & -0.015 & -0.25 & -0.33 
& -0.38 & \cr
144287 & G8 V & 0.470 & 0.270 & 0.325 & 0.050 & -0.087 & -0.40 & -0.58 
& -0.09 & \cr 
144579 & G8 V & 0.455 & 0.232 & 0.226 & 0.061 & 0.013 & -0.51 & -0.47 
& -0.68 & \cr 
145809 & G3 V & 0.396 & 0.187 & 0.328 & 0.030 & -0.042 & -0.20 & -0.35 
& -0.20 & \cr 
146233 & G1 V & 0.405 & 0.211 & 0.345 & 0.014 & -0.070 & -0.04 & -0.28 
&  0.01 & 0.02 \cr 
146775 & G0 V & 0.379 & 0.211 & 0.318 & -0.006 & -0.009 & 0.16 & 0.17 
&  0.11 & \cr
147231 & dG5 & 0.443 & 0.217 & 0.328 & 0.056 & -0.085 & -0.46 & -0.62 
& -0.23 & \cr 
147513 & G3/5 V & 0.400 & 0.198 & 0.324 & 0.023 & -0.043 & -0.13 & -0.29 
& -0.12 & \cr 
147584 & G0 V & 0.352 & 0.177 & 0.323 & 0.014 & 0.020 & -0.04 & -0.11 
& -0.06 & -0.15 \cr 
149414 & G5 Ve & 0.472 & 0.202 & 0.155 & 0.122 & 0.083 & -1.12 & -0.80 
& -1.40 & -1.14 \cr 
150433 & dG2 & 0.410 & 0.173 & 0.313 & 0.057 & -0.044 & -0.47 & -0.55 
& -0.47 & \cr 
150474 & G8 V & 0.477 & 0.261 & 0.364 & 0.073 & -0.125 & -0.63 & -0.83 
& -0.04 & \cr 
150706 & dG3 & 0.389 & 0.188 & 0.312 & 0.024 & -0.017 & -0.14 & -0.25 
& -0.18 & \cr 
152391 & G8 V & 0.456 & 0.285 & 0.298 & 0.009 & -0.059 & 0.01 & -0.22 
& -0.06 & \cr 
153631 & G2 V & 0.386 & 0.189 & 0.339 & 0.020 & -0.040 & -0.10 & -0.27 
& -0.09 & \cr
154088 & G8 IV-V & 0.514 & 0.306 & 0.425 & 0.112 & -0.184 & -1.02 & -1.23 
&  0.10 & \cr
154345 & G8 V & 0.452 & 0.261 & 0.285 & 0.027 & -0.045 & -0.17 & -0.33 
& -0.20 & \cr
155918 & G2 V & 0.391 & 0.144 & 0.277 & 0.069 & 0.016 & -0.59 & -0.53 
& -0.75 & \cr
156274 & G8 V & 0.482 & 0.291 & 0.259 & 0.054 & -0.020& -0.44 & -0.48 
& -0.40 & \cr
156365 & G3 V & 0.418 & 0.230 & 0.406 & 0.009 & -0.146 & 0.01 & -0.39 
&  0.20 & \cr
156826 & G9 V & 0.519 & 0.296 & 0.327 & 0.134 & -0.087 & -1.24 & -1.21 
& -0.25 & \cr
157214 & G2 V & 0.409 & 0.182 & 0.309 & 0.047 & -0.039 & -0.37 & -0.47 
& -0.38 & -0.41 \cr
159222 & G5 V & 0.408 & 0.212 & 0.365 & 0.016 & -0.094 & -0.06 & -0.34 
&  0.04 & \cr
159656 & G5 V & 0.412 & 0.199 & 0.352 & 0.034 & -0.085 & -0.24 & -0.45 
& -0.13 & \cr
159704 & G8 V & 0.469 & 0.269 & 0.322 & 0.049 & -0.084 & -0.39 & -0.57 
& -0.10 & \cr
159868 & G5 V & 0.451 & 0.219 & 0.356 & 0.067 & -0.116 & -0.57 & -0.76 
& -0.20 & \cr
160269 & G0 Va & 0.399 & 0.184 & 0.327 & 0.036 & -0.045 & -0.26 & -0.39 
& -0.25 & \cr
160691 & G5 V & 0.432 & 0.244 & 0.395 & 0.013 & -0.146 & -0.03 & -0.41 
&  0.20 & 0.16 \cr
161612 & G8 V & 0.447 & 0.246 & 0.375 & 0.033 & -0.134 & -0.23 & -0.54 
&  0.08 & \cr
163840 & G2 V & 0.406 & 0.220 & 0.357 & 0.006 & -0.084 & 0.04 & -0.25 
&  0.11 & \cr
165185 & G5 V & 0.388 & 0.180 & 0.315 & 0.031 & -0.018 & -0.21 & -0.31 
& -0.25 & \cr
165401 & G2 V & 0.393 & 0.164 & 0.287 & 0.051 & 0.003 & -0.41 & -0.42 
& -0.51 & -0.46 \cr
168009 & G2 V & 0.411 & 0.199 & 0.351 & 0.032 & -0.083 & -0.22 & -0.44 
& -0.13 & \cr
168060 & G5 V & 0.463 & 0.267 & 0.396 & 0.040 & -0.158 & -0.30 & -0.64 
&  0.16 & \cr
168443 & G8 V & 0.455 & 0.233 & 0.377 & 0.060 & -0.138 & -0.50 & -0.75 
& -0.06 & \cr
171067 & G8 V & 0.424 & 0.234 & 0.309 & 0.012 & -0.054 & -0.02 & -0.24 
& -0.05 & \cr
171665 & G5 V & 0.420 & 0.230 & 0.309 & 0.012 & -0.050 & -0.02 & -0.22 
& -0.05 & \cr
172051 & G5 V & 0.418 & 0.216 & 0.277 & 0.023 & -0.017 & -0.13 & -0.25 
& -0.26 & \cr
175541 & G8 V & 0.564 & 0.294 & 0.413 & 0.248 & -0.197 & -2.38 & -2.27 
& -0.26 & \cr
176377 & G2 V & 0.385 & 0.181 & 0.291 & 0.028 & 0.010 & -0.18 & -0.23 
& -0.28 & \cr
176982 & dG5 & 0.463 & 0.224 & 0.348 & 0.083 & -0.110 & -0.73 & -0.87 
& -0.25 & \cr}}
\endtable
\begintable*{5}
\caption{{\bf Table 1} -- {\it continued}}
{\halign{%
\hfil \rm # \hfil & \quad \rm \hfil # \hfil & \quad $#$ \hfil 
& \quad $#$ \hfil & \quad $#$ \hfil & \quad \hfil $#$
 & \quad \hfil$#$ & \quad \hfil$#$ & \quad \hfil$#$ & \quad \hfil$#$
& \quad \hfil$#$ \cr
HD \ & MK type & b-y & \ \ m_1 & \ \ c_1 & \delta m_1 & \delta c_1\  &
{\rm [Fe/H]}_1 & {\rm [Fe/H]}_2 & {\rm [Fe/H]}_3& {\rm [Fe/H]}_{\rm S} \cr
\noalign{\vskip 10 pt}
178428 & G5 V & 0.438 & 0.246 & 0.354 & 0.020 & -0.109 & -0.10 & -0.39 
&  0.08 & \cr
181321 & G5 V & 0.400 & 0.192 & 0.302 & 0.029 & -0.021 & -0.19 & -0.30 
& -0.24 & \cr
181655 & G8 V & 0.420 & 0.234 & 0.322 & 0.008 & -0.063 & 0.02 & -0.22 
&  0.02 & \cr
185454 & G5 V & 0.427 & 0.273 & 0.309 & -0.023 & -0.056 & 0.33 & 0.02 
&  0.12 & \cr
186408 & G2 V & 0.410 & 0.214 & 0.375 & 0.016 & -0.106 & -0.06 & -0.36 
&  0.06 & 0.21 \cr
186427 & G5 V & 0.416 & 0.226 & 0.354 & 0.011 & -0.092 & -0.01 & -0.30 
&  0.09 & 0.10 \cr
187923 & G0 V & 0.415 & 0.192 & 0.342 & 0.044 & -0.079 & -0.34 & -0.52 
& -0.24 & 0.06 \cr
189340 & G0 V & 0.371 & 0.194 & 0.328 & 0.006 & -0.008 & 0.04 & -0.10 
&  0.03 & \cr
189567 & G2 V & 0.406 & 0.185 & 0.297 & 0.041 & -0.024 & -0.31 & -0.40 
& -0.36 & -0.30 \cr
190067 & G8 V & 0.452 & 0.233 & 0.287 & 0.055 & -0.047 & -0.45 & -0.54 
& -0.33 & \cr
190248 & G8 V & 0.458 & 0.310 & 0.366 & -0.012 & -0.127 & 0.22 & -0.19 
&  0.27 & 0.30 \cr
190360 & G8 IV-V & 0.461 & 0.275 & 0.372 & 0.028 & -0.134 & -0.18 & -0.51 
&  0.15 & 0.26 \cr
190406 & G1 V & 0.387 & 0.187 & 0.347 & 0.023 & -0.049 & -0.13 & -0.31 
& -0.10 & \cr
192020 & G8 V & 0.500 & 0.393 & 0.280 & -0.009 & -0.039 & 0.19 & 0.24 
& -0.19 & \cr
193664 & G5 V & 0.379 & 0.176 & 0.330 & 0.029 & -0.021 & -0.19 & -0.30 
& -0.20 & 0.06 \cr
194640 & G5 V & 0.440 & 0.274 & 0.301 & -0.006 & -0.057 & 0.16 & -0.11 
&  0.01 & \cr
195987 & G9 V & 0.479 & 0.292 & 0.271 & 0.046 & -0.032 & -0.36 & -0.45 
& -0.31 & \cr
196761 & G8 V & 0.440 & 0.251 & 0.269 & 0.017 & -0.025 & -0.07 & -0.22 
& -0.25 & \cr
196850 & G2 V & 0.396 & 0.194 & 0.328 & 0.023 & -0.042 & -0.13 & -0.29 
& -0.13 & \cr
197076 & G5 V & 0.397 & 0.189 & 0.325 & 0.029 & -0.040 & -0.19 & -0.33 
& -0.19 & \cr
197214 & G5 V & 0.423 & 0.217 & 0.250 & 0.028 & 0.006 & -0.18 & -0.24 
& -0.42 & \cr
199288 & G0 V & 0.385 & 0.146 & 0.266 & 0.063 & 0.035 & -0.53 & -0.44 
& -0.71 & \cr
202457 & G5 V & 0.432 & 0.221 & 0.368 & 0.036 & -0.119 & -0.26 & -0.54 
& -0.03 & \cr
202573 & G5 V & 0.567 & 0.237 & 0.479 & 0.312 & -0.266 & -3.02 & -2.89 
& -0.70 & \cr
202628 & G5 V & 0.404 & 0.198 & 0.319 & 0.026 & -0.043 & -0.16 & -0.32 
& -0.16 & \cr
202940 & G5 V & 0.450 & 0.247 & 0.273 & 0.037 & -0.033 & -0.27 & -0.38 
& -0.31 & \cr
203244 & G5 V & 0.444 & 0.250 & 0.264 & 0.025 & -0.022 & -0.15 & -0.27 
& -0.30 & \cr
205905 & G4 IV-V & 0.388 & 0.219 & 0.329 & -0.008 & -0.032 & 0.18 & 0.22 
&  0.15 & \cr
206827 & G2 V & 0.241 & 0.148 & 0.520 & -0.029 & 0.967 & *** & *** 
& *** & \cr
206860 & G0 V & 0.376 & 0.186 & 0.318 & 0.017 & -0.005 & -0.07 & -0.18 
& -0.10 & \cr
207129 & G2 V & 0.386 & 0.181 & 0.340 & 0.028 & -0.041 & -0.18 & -0.33 
& -0.18 & \cr
210460 & G0 V & 0.451 & 0.208 & 0.328 & 0.078 & -0.088 & -0.68 & -0.79 
& -0.36 & \cr
210918 & G5 V & 0.415 & 0.195 & 0.323 & 0.041 & -0.060 & -0.31 & -0.46 
& -0.25 & \cr
211038 & G8 V & 0.556 & 0.290 & 0.433 & 0.232 & -0.210 & -2.22 & -2.18 
& -0.21 & \cr
211415 & G1 V & 0.386 & 0.187 & 0.292 & 0.022 & 0.007 & -0.12 & -0.19 
& -0.22 & \cr
211998 & G0 V & 0.447 & 0.116 & 0.240 & 0.163 & 0.001 & -1.53 & -1.26 
& -1.43 & -1.61 \cr
212330 & G4 V & 0.424 & 0.206 & 0.364 & 0.040 & -0.109 & -0.30 & -0.55 
& -0.12 & -0.17 \cr
212697 & G3 V & 0.392 & 0.204 & 0.284 & 0.010 & 0.007 & 0.00& -0.10 
& -0.15 & 0.30 \cr
213628 & G3 V & 0.441 & 0.266 & 0.315 & 0.004 & -0.071 & 0.06 & -0.21 
&  0.03 & \cr
213941 & G5 V & 0.416 & 0.196 & 0.283 & 0.041 & -0.021 & -0.31 & -0.39 
& -0.37 & \cr
214615 & G8/K0 V & 0.469 & 0.306 & 0.301 & 0.012 & -0.063 & -0.02 & -0.25 
& -0.06 & \cr
214953 & G1 V & 0.362 & 0.171 & 0.379 & 0.025 & -0.048 & -0.15 & -0.32 
& -0.21 & \cr
217014 & G4 V & 0.416 & 0.232 & 0.364 & 0.005 & -0.102 & 0.05 & -0.27 
&  0.16 & 0.06 \cr
218209 & G6 V & 0.419 & 0.189 & 0.258 & 0.051 & 0.001 & -0.41 & -0.42 
& -0.54 & \cr
218640 & G2 V & 0.418 & 0.209 & 0.744 & 0.030 & -0.484 & *** & *** 
& *** & \cr
219048 & G5 V & 0.457 & 0.229 & 0.329 & 0.067 & -0.090 & -0.57 & -0.72 
& -0.23 & \cr
219709 & G2 V & 0.395 & 0.210 & 0.336 & 0.006 & -0.049 & 0.04 & -0.18 
&  0.05 & \cr
221818 & G8 V & 0.469 & 0.297 & 0.286 & 0.021 & -0.048 & -0.11 & -0.29 
& -0.16 & \cr
223498 & G7 V & 0.456 & 0.272 & 0.366 & 0.022 & -0.127 & -0.12 & -0.45 
&  0.15 & \cr
224465 & dG2 & 0.415 & 0.224 & 0.338 & 0.012 & -0.075 & -0.02 & -0.27 
&  0.04 & \cr
224930 & G3 V & 0.428 & 0.189 & 0.215 & 0.063 & 0.037 & -0.53 & -0.44 
& -0.80 & -0.70 \cr
225239 & G2 V & 0.412 & 0.169 & 0.312 & 0.064 & -0.045 & -0.54 & -0.60 
& -0.53 & -0.50 \cr
225261 & G9 V & 0.453 & 0.271 & 0.252 & 0.018 & -0.012 & -0.08 & -0.20 
& -0.34 & \cr}}
\endtable

\subsection{Halo contamination}

\tx Our sample may include some halo stars. To exclude them, we introduced 
a {\it chemical} criterion, according to which all stars having 
[Fe/H] $< -1.2$ are considered to be halo members. The application of this 
condition to the objects of Table 1 excludes four stars, so that our final 
sample contains 287 objects. 

\beginfigure{2}
\vskip 8.5 cm
\caption{{\bf Figure 2.} \rm Metallicity distribution of 287 
    dwarf stars with spectral types in the range G0 -- G9 (continuous 
    line), and 231 dwarfs of spectral types G2 -- G9 (squares).}
\endfigure
\begintable*{6}
\caption{{\bf Table 2.} Abundance distribution in the 
solar neighbourhood.}
{\halign{%
\rm \hfil# \hfil & \quad \hfil # \hfil & \quad \hfil# \hfil 
& \quad \hfil#  & \quad \hfil# \hfil & \quad \hfil #
 & \quad \hfil#\hfil & \quad \hfil#\hfil\cr
[Fe/H] & $\phi$ & $\Delta N_0$ & $\delta(\Delta N)_0$ & $f$ 
& $\delta(\Delta N)_1$ & 
log$\biggl[{\Delta N_0+\delta(\Delta N)_0\over\Delta\phi}\biggr]$ &
log$\biggl[{\Delta N_0/f+\delta(\Delta N)_1\over\Delta\phi}\biggr]$
\cr
\noalign{\vskip 10 pt}
$-$1.2 to $-$1.1 & 0.25 to 0.28 & 0 & 0.00 & 0.23  & $-$0.06 & 
&  \cr
 & & & & & & & \cr
$-$1.1 to $-$1.0 & 0.28 to 0.32 & 0 & $-$0.03 & 0.23  & $-$0.17 &
 & \cr
 & & & & & & $1.22(+0.21,-0.43)$ & $1.89(+0.11,-0.15)$ \cr
$-$1.0 to $-$0.9 & 0.32 to 0.35 & 0 & $-$0.11 & 0.23  & $-$0.40 & 
& \cr
 & & & & & & & \cr
$-$0.9 to $-$0.8 & 0.35 to 0.40 & 3 & $-$0.36 & 0.23  & $-$0.80 & 
&  \cr
 & & & & & & & \cr
$-$0.8 to $-$0.7 & 0.40 to 0.45 & 6 & $-$0.92 & 0.36  & $-$1.27 & 
$2.01(+0.16,-0.25)$ & $2.49(+0.10,-0.13)$  \cr
 & & & & & & & \cr
$-$0.7 to $-$0.6 & 0.45 to 0.50 & 8 & $-$1.78 & 0.53  & $-$1.54 &
$2.09(+0.15,-0.22)$ & $2.43(+0.10,-0.14)$\cr
 & & & & & & & \cr
$-$0.6 to $-$0.5 & 0.50 to 0.56 & 15 & $-$2.37 & 0.79  & $-$1.17 & 
$2.32(+0.11,-0.14)$ & $2.47(+0.09,-0.12)$\cr
 & & & & & & & \cr
$-$0.5 to $-$0.4 & 0.56 to 0.63 & 21 & $-$1.61   & 0.85  & 0.07 & 
$2.44(+0.09,-0.11)$ & $2.55(+0.08,-0.10)$\cr
 & & & & & & & \cr
$-$0.4 to $-$0.3 & 0.63 to 0.71 & 27 & 1.01   & 0.98  & 1.82 & 
$2.54(+0.08,-0.09)$ & $2.56(+0.07,-0.09)$\cr
 & & & & & & & \cr
$-$0.3 to $-$0.2 & 0.71 to 0.79 & 56 & 4.14   & 0.99  & 3.19 & 
$2.88(+0.05,-0.06)$ & $2.87(+0.05,-0.06)$\cr
 & & & & & & & \cr
$-$0.2 to $-$0.1 & 0.79 to 0.89 & 45 & 5.31   & 1.00  & 3.33 & 
$2.70(+0.06,-0.07)$ & $2.68(+0.06,-0.07)$\cr
 & & & & & & & \cr
$-$0.1 to 0.0  & 0.89 to 1.00 & 41 & 3.45     & 1.00  & 2.16 & 
$2.61(+0.06,-0.07)$ & $2.59(+0.06,-0.07)$\cr
 & & & & & & & \cr
0.0 to 0.1 & 1.00 to 1.12 & 36 & 0.22         & 1.00  & 0.40 & 
$2.48(+0.07,-0.08)$ & $2.48(+0.07,-0.08)$\cr
 & & & & & & & \cr
0.1 to 0.2 & 1.12 to 1.26 & 22 & $-$1.98 & 1.00          & $-$0.99 & 
$2.16(+0.09,-0.11)$ & $2.18(+0.09,-0.11)$\cr
 & & & & & & & \cr
0.2 to 0.3 & 1.26 to 1.41 & 6 & $-$2.30 & 1.00        & $-$1.53 & 
 & \cr
 & & & & & & $0.99(+0.19,-0.36)$ & $1.11(+0.17,-0.30)$ \cr
0.3 to 0.4 & 1.41 to 1.58 & 1 & $-$1.56 & 1.00        & $-$1.36 &
 & \cr}}
\endtable
A {\it kinematical} criterion was also considered, namely, objects having 
radial velocities $\vert v_{\rm r}\vert$ or tangential velocities 
$\vert v_{\rm t}\vert $ larger than 100 km s$^{-1}$ are likely to be halo members. 
This criterion tends to remove objects with large velocity components, a 
characteristic generally shared by halo stars. However, it should be 
recalled that our sample comprises stars within 25 pc from the Sun only. 
Since parallaxes are preferentially measured in stars with large proper 
motions, large velocity components could be obtained even for disc stars, 
owing to their proximity. In fact, using radial velocities from Gliese \& 
Jahreiss (1991), and calculating tangential velocities from proper motions 
and parallaxes of the same source, we find that 19 stars should be excluded 
from our initial sample, according to the kinematical criterion. The four 
objects excluded on the basis of the chemical criterion also belong to 
this group, which reinforces the suspicion that they are halo members. 
The remaining 15 objects have disc metallicities. Since the resulting 
metallicity distribution is not affected by the kinematical criterion,   
we have decided to take into account the chemical criterion only. 

\subsection{Metal content and lifetimes of the hotter G dwarfs}

\tx The relative metallicity distribution of our adopted sample is shown in 
Fig. 2 (continuous line), and the corresponding data are given in Table 2. 
In this table, column 1 gives the adopted metallicity intervals, and 
column 3 gives the raw absolute distribution, which we will call 
$\Delta N_0$. 

A remark should be made with respect to the spectral types considered. Pagel 
\& Patchett (1975) used only stars with spectral types between G2 V and 
G8 V. Earlier stars, with spectral types G0 V and G1 V, were not used, 
since their lifetimes are considered as slightly lower than the assumed age 
of the Galaxy. On the other hand, G9 V stars were not included, as their low 
effective temperatures invalidate their adopted calibration. The latter 
problem does not occur with our adopted calibrations, so that we have 
included G9 V stars in our sample. 

Data selection based on a given spectral range has some limitations, in the 
sense that the spectral types depend both on the effective temperature 
and on the metal content. Moreover, the stellar chemical composition also 
affects its lifetime.

The effect of differing metal content would be a modification of the 
mass range of the stars classified as G dwarfs, which would affect the 
homogeneity of our sample. In this case, the G dwarf metallicity 
distribution would differ from the corresponding distribution of the 
long-lived stars for which the theoretical results apply. On the other 
hand, the effect of the chemical composition on the lifetimes could also 
produce a `G dwarf problem' (Bazan \& Mathews 1990; Meusinger \& 
Stecklum 1992), as metal-poor stars have lower lifetimes, and some 
of them may have already died. However, the magnitudes of these effects 
are not at all clear, partly because they are strongly dependent on 
stellar evolution models. 

In order to check the effects of the hotter stars, we plotted in Fig.~2  
the metallicity distribution of the 231 objects with spectral types 
between G2 and G9 V (squares). As can be seen from the figure, the plotted 
distributions are very similar, suggesting that the inclusion of the 
earlier stars does not imply a biased sample. This conclusion is 
reinforced by Fig. 3, where we plot the metallicity of each star (dots), 
along with the mean value for each spectral type (squares). Here, G8 and G9 
stars are merged in the same bin. Since G0 and G1 V stars are hotter and 
have shorter lifetimes, we would expect their metallicity to be larger 
than for the remaining, cooler stars. We see that there is a good 
agreement of the metallicity of G0 and G1 stars with the average values 
of the remaining spectral types. Also, the scatter around the mean value 
is similar for all spectral types, except types G6 and G7, for which 
there are fewer stars. Therefore, we believe that there is no reason to 
exclude such objects from our sample, which comprises then 287 stars with 
spectral types in the range G0 -- G9 V.
\beginfigure{3}
\vskip 5.5 cm
\caption{{\bf Figure 3.} Abundances as a function of the 
    spectral type (dots), and mean values for each type (squares). 
    Stars of types G8 and G9 are merged in the same bin.}
\endtable
\section{Additional Corrections}

\tx Following Pagel \& Patchett (1975) and Pagel (1989), we have corrected 
the final distribution for observational errors and cosmic scater 
in the age--metallicity relation, assuming a standard deviation of 0.1 dex. 
For purposes of deconvolution, the raw distribution in column~3 of Table~1 
was fitted to a Gaussian given by $(A/\sigma\sqrt {2\pi}) 
\ {\rm exp} [-({\rm [Fe/H]}-\mu)^2/2\sigma^2]$, where $A = 28.48, 
\mu = -0.16$, and $\sigma = 0.23$. The deconvolved Gaussian has $\sigma 
= 0.21$; the corrections $\delta(\Delta N)_0$ are obtained by the  
subtraction of the first Gaussian from its deconvolved version, and are 
given in the fourth column of Table~2. We have also considered a 
standard deviation $\sigma = 0.15$ dex, which is of the order of the error 
estimated for calibration~1. In this case, the final distribution does 
not appreciably change in comparison with the results presented in Table~2. 

We have also made use of the oxygen abundance rather than the iron 
abundance, in order to be consistent with chemical evolution models that 
use the instantaneous recycling approximation. We adopt the same [O/Fe] 
ratio as given by Pagel (1989): 
$$\log \phi = {\rm [O/H]} = 0.5{\rm [Fe/H]}. \eqno\stepeq$$
The oxygen abundances $\phi$ are given in the second column of Table~2. 

As in Pagel (1989), our distribution is spatially limited within 25 pc of the 
Sun, which may introduce a non-negligible bias in the metallicity 
distribution, as was shown by Sommer-Larsen (1991). Since older stars 
generally have lower metallicities and larger scale heights relative to 
the galactic plane, we expect their relative number to be artificially 
reduced by the limitation of our sample within 25 pc of the Sun. To solve this 
problem, we have adopted the correction procedure introduced by Sommer-Larsen 
(1991), who defined a weight factor $f$, which is given in column~5 of 
Table~2. This factor has been determined by Sommer-Larsen (1991) on the 
basis of kinematical models by Norris \& Ryan (1989) and Kuijken \& 
Gilmore (1989). According to Sommer-Larsen (1991), $f$ is essentially 
independent of the specific model parameters, so that we adopted the 
values obtained for the models by Kuijken \& Gilmore (1989) which reproduce 
the rotation curve and surface density in a maximum disc type model (cf. 
Sellwood \& Sanders 1988). Following Sommer-Larsen (1991), we 
have applied the correction factors $f^{-1}$ to the {\it raw data} 
as given in column~3 of Table~2, and again fitted a Gaussian to the 
resulting distribution, for which we obtained $A = 31.95, \mu = -0.19,$ and 
$\sigma = 0.27$. This was also deconvolved to correct for observational 
errors and cosmic scatter assuming a deviation of 0.1. The calculated 
deviation becomes $\sigma = 0.26$, and the new correction factors, which 
we denote as $\delta(\Delta N)_1$, are determined as before, and 
given in  column~6 of Table~2.

The obtained relative distributions are given in the last two columns of 
Table~2. Column~7 gives the raw distribution corrected for observational 
errors and cosmic scatter {\it only}, and  column~8 gives the final 
distribution, which also includes the scale height correction according 
to Sommer-Larsen (1991). In these columns, the first four bins were grouped 
to yield a mean value for the distribution. This procedure was also used 
for the last two bins.

\section{Consequences for chemical evolution}

\beginfigure{4}
 \vskip 8.5 cm
 \caption{{\bf Figure 4.} A comparison of our raw metallicity distribution 
  (thick continuous line) and the corresponding quantities from Pagel (1989, 
   broken line) and Rana \& Basu (1990, thin line).}
 \endfigure
\tx Fig. 4 shows our raw relative distribution  $\Delta  N_0 / 
287$ (thick continuous line) as a function of the metallicity in comparison 
with the equivalent quantity from Pagel, $\Delta N / 132$ (broken 
line, cf. Pagel 1989, p. 210). Also shown is the metallicity distribution 
of Rana \& Basu (1990, thin line). It can be  seen that our distribution 
presents a single, main peak located between the main peaks of the 
previously quoted distributions. Our distribution is narrower than the one 
by Pagel (1989), and has a prominent peak around $-$0.20 dex, and is in 
this respect similar to the distribution by Rana \& Basu (1990). However, 
our distribution does not confirm the large number of stars with 
${\rm [Fe/H]} > 0.1$ dex found by these authors. Instead, it shows a small 
percentage of these stars, similar to that found by Pagel. 

A remarkable feature of our distribution is that it shows no stars in 
the interval $-1.2 \leq {\rm [Fe/H]} \leq -0.8$. This does not mean 
that such stars do not exist (a metallicity distribution based on 
calibration 1, for instance, shows 12 stars in this range), but that 
their number is small. This result not only confirms the G dwarf problem, 
but aggravates it, in comparison with previous distributions at this 
metallicity range.

Another feature of our distribution is the strong peak observed, which 
could be interpreted as the result of a star formation burst. To  
illustrate this, let us assume that the age of a star can be determined 
on the basis of an age--metallicity relation (AMR). Of course, there 
is a large scatter in this relation (cf. Edvardsson et al. 1993), so 
that this should be considered for illustration purposes only. 
Adopting the parametrization of the AMR given by Rana (1991), we have 
$${\rm [Fe/H]} = 0.68 - {11.2 \over {8+t}}, \eqno\stepeq$$
where $t$ is the age, taken as $t = 0$ at the birth of the disc, and 
$ t = 12 $ Gyr for the disc age. Assuming that the burst occurred in the 
range  $-0.10 \leq {\rm [Fe/H]} \leq -0.30$, it can be concluded that it 
occurred some 5 to 8 Gyr ago, similar to the results found by other 
authors on the basis of different methods (cf. Noh \& Scalo 1990, and 
references therein).

\beginfigure{5}
\vskip 6.5 cm
\caption{{\bf Figure 5.}  An adaptation of Fig. 5 by 
    Pagel (1989) with the inclusion of results of the present work. 
    Filled circles: corrected sample (Table 2, column 8); 
    empty circles: partially corrected sample (Table 2, column 7); 
    crosses: results by Pagel (1989); continuous lines: 
    model fits as given by Pagel (1989).} 
\endfigure

\beginfigure{6}
\vskip 8.5 cm
\caption{{\bf Figure 6.} Raw metallicity distribution of G 
    dwarfs (thick continuous line) compared with two chemical evolution 
    models from Giovagnoli \& Tosi (1995), corresponding to their standard 
    model (broken line) and model 3 (thin line).} 
 \endfigure   

Fig. 5 is based on Fig. 5 of Pagel (1989). Apart 
from Pagel's results (crosses), we include (i) our results for the sample 
corrected for observational errors and cosmic scatter {\it only} (column 
7 of Table 2, empty circles), and (ii) the results corrected also for 
the scale height effect (column 8 of Table 2, filled circles). Both 
distributions were also normalized for the total number of stars (132 stars) 
in Pagel's sample, in order to make the qualitative differences more 
evident. This corresponds to multiplying the terms within brackets of columns 
7 and 8 of Table 2 by 0.46 and 0.41, respectively. Our error bars are 
similar to those in Pagel's work, and have been omitted from Fig.~5 for 
the sake of clarity. 
The figure also shows the predictions of (iii) the simple model with 
effective yield $p = 0.5\ {\rm Z}_\odot$, and present gas fraction $\mu_1 
= 0.1$; (iv) the simple model with initial production spike, or prompt 
initial enrichment (PIE); (v) Lynden-Bell's (1975) Best Accretion Model 
with parameters $M = 20$, $p = {\rm Z}_\odot$, and $\mu_1 = 0.2$; (vi) and 
(vii) Clayton's (1985, 1988) standard infall models with parameters $k = 2$, 
$M = 18.5$, $p = 0.75\ {\rm Z}_\odot$, $\mu_1 = 0.12$, and $k = 4$, $M = 65$, 
$p = 0.75\ {\rm Z}_\odot$, and $\mu_1 = 0.12$, respectively. Models 
(iii)--(vii) are shown as continuous lines labelled in the figure. The 
analytical formulae and a discussion of the model 
parameters are given by Pagel (1989). As can be seen from the figure, the 
simple model does not fit the distribution very well, even at large 
abundances. The use of different values for the effective yield does not 
help this situation. The simple model with an effective yield does not 
seem to be a good approximation to the chemical evolution of the solar 
neighbourhood. It can also be seen that infall models tend to decrease 
the differences between the theoretical and observed distributions.

The  G dwarf problem is evident in this plot. The simple model predicts 
too many metal-poor stars relative to what is found in our vicinity. 
However, the G dwarf problem may not only be the paucity of metal-poor 
dwarfs. A smaller number of metal-rich dwarfs is also found relative to 
the simple model predictions, as shown in Fig.~5. This feature is 
apparent in some previous metallicity distributions (e.g. Pagel 1989), but 
has not been commented upon. The use of a larger star sample, preferably 
having metallicities obtained by high-resolution spectroscopy, may in the 
future confirm whether or not this feature is real.

The application of the Sommer-Larsen correction to our sample increases 
the number of metal-poor stars, and alleviates somewhat the G dwarf 
problem, but does not completely solve it. On the other hand, the  
agreement between the data and the model curves, especially Clayton's 
standard model with $k = 4$, and Lynden-Bell's model, improves very much. 

Fig. 6 shows a comparison between our raw metallicity distribution 
(thick continuous line) and two infall models from the recent work of 
Giovagnoli \& Tosi (1995), corresponding  to their standard model (broken 
line) and model~3 (thin line). Both models use an exponentially decreasing 
star formation rate with e-folding time $\tau=15$ Gyr,  and a constant infall 
rate of $4 \times 10^{-3}\ {\rm M}_\odot\ {\rm kpc}^{-2}\ {\rm yr}^{-1}$. 
The main difference between the models is that the standard model uses 
classic yields from the literature (D\'{\i}az \& Tosi 1986; Tosi 1988), 
and Tinsley's (1980) initial mass function, while model~3 uses 
metallicity-dependent yields given by Maeder (1992), and Salpeter's (1955) 
initial mass function. The agreement between 
the data and the models is good, but they still predict an excess 
of metal-poor stars. The problem could be with a constant infall rate. An 
infall like that of Lynden-Bell, which rises to a maximum 
during the lifetime of the Galaxy, and has a smooth decay  later 
on, may produce the small number of metal-poor stars observed, as suggested 
by Fig.~5. Another possible reason for the observed discrepancy at 
lower metallicities is that Giovagnoli \& Tosi (1995) treat the disc as 
formed out of metal-free gas, and do not take into account the effect of  
the halo or the thick disc. If these effects were considered, it could be 
expected that the disc initial metallicity would be different from zero, 
and thus fewer metal-poor stars would be produced (Tosi 1995, private 
communication).

\section*{Acknowledgments}

\tx We thank Gustavo F. Porto de Mello and Cristina Chiappini for some 
helpful discussions, and Dr B.E.J. Pagel for his comments on an earlier 
version of this paper. This work was partially supported by CNPq and FAPESP. 

\section*{Note added in proof}

\tx After this paper has been submitted for publication, we learned about 
another determination of the metallicity distribution of G dwarfs 
by R. F. G. Wyse and G. Gilmore (1995, preprint). Their distribution 
comprises 128 F and G dwarfs from the catalogue of Gliese \& 
Jahreiss (1991). The stellar metallicities are obtained by the use of the 
calibration of Schuster \& Nissen (1989), also presented in this paper, and 
photometric data are taken from Olsen (1983). The comparison between their 
original metallicity distribution (not corrected for the inclusion of stars with 
lifetime shorter than the Galactic age) with our raw metallicity distribution 
shows a remarkable agreement.

\end